\documentclass[aps,showpacs,floatfix,twocolumn,amsmath,amssymb]{revtex4}
\usepackage{graphicx}
\usepackage{epsfig}
\usepackage{lscape}
\usepackage{ulem}
\usepackage{morefloats}
\usepackage{rotating}
\usepackage{color}

\begin{document}

\title{Phase transitions in Core-Collapse Supernova Matter at sub-saturation densities}

\author{Helena Pais$^{1,2}$, William G. Newton$^3$, and Jirina R. Stone$^{2,4}$}
\affiliation{$^1$Centro de F\'isica Computacional, Department of Physics, University of Coimbra, P-3004-516 Coimbra, Portugal \\
$^2$Department of Physics and Astronomy, University of Tennessee, Knoxville, TN 37996, USA \\
$^3$Department of Physics and Astronomy, Texas A\&M University - Commerce PO Box 3011, Commerce, TX 75429-3011, USA \\
$^4$Department of Physics, University of Oxford, Oxford OX1 3PU,  United Kingdom}

\begin{abstract}

Phase transitions in hot, dense matter in the collapsing core of massive stars have an important impact on the core-collapse supernova mechanism as they absorb heat, disrupt homology and so weaken the developing shock. We perform a three-dimensional, finite temperature Skyrme-Hartree-Fock (SHF) study of inhomogeneous nuclear matter to determine the critical density and temperature for the phase transition between the pasta phase and homogeneous matter and its properties. We employ four different parametrizations of the Skyrme nuclear energy-density functional, SkM$^*$, SLy4, NRAPR and SQMC700, which span a range of saturation-density symmetry energy behaviours constrained by a variety of nuclear experimental probes. For each of these interactions we calculate free energy, pressure, entropy and chemical potentials in the range of particle number densities where the nuclear pasta phases are expected to exist, 0.02 - 0.12 fm$^{-3}$, temperatures 2 - 8 MeV and a proton fraction of 0.3. We find unambiguous evidence for a first-order phase transition to uniform matter, unsoftened by the presence of the pasta phases. No conclusive signs of first-order phase transition between the pasta phases is observed, and it is argued that the thermodynamic quantities vary continuously right up to the first-order phase transition to uniform matter. We compare our results with thermodynamic spinodals calculated using the same Skyrme parameterizations, finding that the effect of short-range Coulomb correlations and quantum shell effects included in our model leads to the pasta phases existing at densities up to 0.1 fm$^{-3}$ above the spinodal boundaries, thus increasing the transition density to uniform matter by the same amount. The transition density is otherwise shown to be insensitive to the symmetry energy at saturation density within the range constrained by the concordance of a variety of experimental constraints, and can be taken to be a well-determined quantity.
\end{abstract}


\maketitle

\section{Introduction} \label{I}

The modeling of core-collapse supernovae (CCSN) encompasses a large variety of physics, from the macroscopic description of the gravitational collapse to the microscopic properties of atomic and subatomic particles. One of the key physical inputs bridging the macro- and micro-physics in CCSN  simulations is the equation-of-state (EoS), connecting the pressure of stellar matter to its energy density and temperature and composition, determined by the underlying interactions between the constituent particles. 

The essential physical ideas of the CCSN scenario are the rebound of the central region of the core upon reaching nuclear matter densities, the development and stalling of a shock wave as the rebounding material encounters material further out, and the revival of the shock wave by neutrino pressure. These processes occur in hot matter (temperatures up to $T\sim20$ MeV) that bridges a density region in which inhomogeneous matter consisting of heavy nuclei, nucleons and light clusters (deuterium, tritium, helions, $\alpha$-particles) transitions to uniform nuclear matter \citep{Brown82, Bethe80, Bethe79}. This transition region is expected to be mediated by heavy quasi-nuclei structures termed nuclear ``pasta'' after their exotic geometries: rods, slabs, cylindrical holes, bubbles and more complex networks of shapes \citep{Ravenhall-83}.  The formation of these phases is driven by the competition between the surface tension and the Coulomb repulsion of closely spaced heavy nuclei, and their phase diagram in density and temperature space has been thoroughly explored phases in the context of both CCSN and neutron star crust matter \citep{Ravenhall-83, Horowitz-04I, Horowitz-05, Maruyama-05, Watanabe-05a, Sonoda-08, Sonoda-10} using a variety of theoretical apparatus. The pasta phases appear in a well established density range $0.01 - 0.1$ fm$^{-3}$ and temperature range $T\lesssim10$MeV.

The equation of state throughout the transition region, and the exact nature of the phase transitions between the different shape phases of nuclear pasta, and from the pasta phases to uniform nuclear matter, all have an important bearing on the evolution of the shock. A stiffer EoS acts against gravity to slow the collapse, causing a greater infall of material from the mantle, increasing the radius at which the shock forms, and weakening the shock.  Any phase transition in the collapsing matter will absorb heat, disrupt the homology, and again weaken the shock. Establishing the nature and strength of these phase transitions is essential to a physically realistic description of CCSN energetics. 

The timescale for CCSN is believed to be of order of seconds and matter does not have enough time to reach $\beta$-equilibrium throughout the rapid changes \cite{Sato-74, Lattimer-91}. Trapped neutrinos have the effect of freezing the lepton fraction, and simulations show that this leads to proton fractions that are approximately constant throughout the imploding core at $y_p \sim 0.3$ \cite{Sato-75, Lattimer-91}.

The nuclear pasta phases can be thought of as arising due to unstable collective modes in uniform matter \citep{ProvidenciaC-06a}, with the transition to uniform matter studies by analyzing the dynamical and thermodynamical spinodals in density and temperature space  \citep{Chomaz-04, Shlomo-04, ProvidenciaC-06, Lucilia-06}. To study the details of the pasta phases themselves, modeling of the inhomogeneous phases is required. A number of early works studying the pasta phases within semiclassical Liquid Drop (LD) and Thomas-Fermi (TF) formalisms predicted that the transitions between pasta phases and the transition to uniform nuclear matter were first order, accompanied by discontinuities in the pressure with increasing density \cite{Lamb83,Williams85,Lattimer-91,Lassaut-87}. In such studies and similar, however, the nuclear pasta shapes expected to appear have to be specified \textit{a priori}, with the ground state phase chosen to be the one that gives the minimum free energy at a particular density. Many recent studies using these method and modern relativistic and non-relativistic nuclear energy density functionals have been conducted \citep{Sonoda-08, Sonoda-10, Avancini-08,Avancini-09,Avancini-10}.

In contrast to the above studies, statistical models in which free nucleons are treated within a mean-field approximation and nuclei are considered to form a loosely interacting cluster gas tend to give a continuous phase transition \cite{Raduta10}, although such models do not take into account the pasta phases.

More recently, the application of quantum and classical molecular dynamics  (QMD, CMD respectively) and three-dimensional Skyrme-Hartree-Fock (3DSHF) approaches have allowed energetically-preferred pasta phases to emerge bias-free during calculations \cite{Maruyama-98,Watanabe-01, Watanabe-05,Horowitz-04I, Horowitz-04II,Dorso-12, Bonche-81, Bonche-82, Magierski-02, Gogelein-08}, resulting in a number of previously unconsidered phases to be studied \citep{Schutrumpf14}. The molecular dynamics methods trade a complete model of the nuclear interactions and quantum shell effects for the modeling of  a very large number of nucleons in a large computational domain. The 3DSHF method self-consistently includes quantum shell effects and a relatively sophisticated nuclear model but is limited to computational cell sizes containing no more than a few thousand nucleons \cite{Newton-09}. 

Uncertainties in the nuclear matter EoS around saturation density are dominated by uncertainties in the nuclear symmetry energy, that component of the nuclear matter binding energy that describes the energy cost decreasing the proton fraction of matter. Much progress has been made in constraining the symmetry energy $S_0$ and its density dependence $L$ around saturation density \citep{Tsang-12, Lattimer13}, with a current concordance of experimental probes of  $S_0 \sim 32$ MeV, $40 \lesssim L \lesssim 60$ MeV. In the light of the most up-to-date set of nuclear constraints on nuclear matter properties in the vicinity of saturation density, coupled with constraints from the maximum mass of neutron stars \citep{Demorest10, Antoniadis13}, Dutra et al. \cite{Dutra-12} tested the capabilities of 240 Skyrme interaction parameter sets, finding that only 5 of these forces satisfied such constraints. Given the sensitivity of the outcomes of CCSN simulations to the nuclear matter EoS the best models of nuclear matter drawn from such studies should be used. 

The work in this paper follows previous work \citep{Newton-09, Newton-phd, Pais-12} in which the pasta phases of CCSN matter were studied using the 3DSHF method. The goal of this work is to use the latest constraints on the Skyrme energy-density functional to  characterize the phase transition between the pasta phases and the uniform phase, where all phases are allowed to emerge self-consistently using a number of thermodynamic quantities. We seek to compare the resulting phase transitions with those that arise from studying the thermodynamic spinodal with the same underlying nuclear interactions, in order to assess the impact of a consistent description of the pasta phases on the density and temperature range of stability of the uniform phase. We selected four different interactions, SkM* \citep{Bartel-82}, SLy4 \citep{Chabanat-98}, NRAPR \citep{Steiner-05} and SQMC700 \citep{Guichon-06}, based on their overall performance in modeling of a wide variety of nuclear matter properties \citep{Dutra-12}. 

In section \ref{II}, we briefly define phase transition and describe its possible characters and in section \ref{III}, the numerical method is explained. In section \ref{IV}, we present and discuss the results obtained and, finally, in section \ref{V}, some conclusions are drawn.

\section{The phase transition} \label{II}

The equilibrium state of a homogeneous body is determined by specifying any two thermodynamic quantities, for example the volume $V$ and the energy $E$. There is, however, no reason to suppose that for every given pair of values of $V$ and $E$ the state of the body corresponding to thermal equilibrium will be homogeneous. It may be that for a given volume and energy in thermal equilibrium the body is not homogeneous, but separates into two homogeneous parts in contact which are in different states. Such states of matter that can exist simultaneously in equilibrium with one another and in contact are described as different phases \citep{Landau80}.

\subsection{First and second order phase transitions}

Phase transitions which are connected with an entropy discontinuity are called discontinuous or phase transitions of first order. On the other hand, phase transitions across which the entropy is continuous are either continuous or of second or higher order.

For a first-order phase transition, at least one of the first derivatives of the free energy with respect to one of its variables is discontinuous \citep{Greiner97}:
$S=-\frac{\partial F}{\partial T}|_{N,V,...}$
$P=-\frac{\partial F}{\partial V}|_{N,T,...}$  
This discontinuity produces a divergence in the higher derivatives like the specific heat $C_V = T\frac{\partial S}{\partial T}|_{V} = - T\frac{\partial^2 F}{\partial^2 T}|_{V}$, or the incompressibility $K(\rho_0)=9\rho_0^2\frac{\partial^2 E_{SNM}(\rho)}{\partial^2\rho}|_{\rho=\rho_0}$, where $E_{SNM}$ is the energy per particle of symmetric nuclear matter \citep{Vidana09}.
  
For a phase transition of second (or $n$th order), the first derivatives of the free energy are continuous; however, second (or $n$th order) derivatives, like the specific heat or the susceptibility, are discontinuous or divergent. The transition to superconductivity without an external magnetic field is an example of phase transitions of this kind \citep{Greiner97}. In this paper we will examine the mediation of the transition between inhomogeneous and homogeneous of nuclear matter at finite temperature by the nuclear pasta phases.

\subsection{Thermodynamical instabilities} \label{IIIa}

In order to analyze the impact of the existence of the pasta phases on the stability of uniform matter, we will calculate the thermodynamical spinodals for the Skyrme interactions used following the method outlined in Ref. \citep{ProvidenciaC-06} and references therein. Matter is stable to fluctuations in density and composition, under conditions of constant volume and temperature, when the free energy density $\cal F$ is a convex function of the proton and neutron densities. These densities are associated with the chemical potentials $\mu_n = \frac{\partial \cal F}{\partial \rho_n}$ and $\mu_p = \frac{\partial \cal F}{\partial \rho_p}$ 
and the free energy curvature is given by
$$
\cal C = \left(\begin{array}{c c}
\dfrac{\partial^2 \cal F}{\partial^2 \rho_p} & \dfrac{\partial^2 \cal F}{\partial \rho_p \partial \rho_n}\\
\dfrac{\partial^2 \cal F}{\partial \rho_n \partial \rho_p} & \dfrac{\partial^2 \cal F}{\partial^2 \rho_n}
\end{array}  \right) = \left(\begin{array}{c c}
\dfrac{\partial \mu_p}{\partial \rho_p} & \dfrac{\partial \mu_n}{\partial \rho_p}\\
\dfrac{\partial \mu_p}{\partial \rho_n} & \dfrac{\partial \mu_n}{\partial \rho_n}
\end{array}  \right)
$$
The eigenvalues of this matrix are given by $$\lambda_{\pm}=\frac{1}{2}(\rm{Tr}(\cal C) \pm \sqrt{\rm{Tr}(\cal C)^{\rm{2}}-\rm{4}\rm{Det}(\cal C)}$$ and the eigenvectors by $$ \frac{\delta \rho_p^{\pm}}{\delta \rho_n^{\pm}} = \dfrac{\lambda_{\pm} - \frac{\partial \mu_n}{\partial \rho_n}}{\frac{\partial \mu_p}{\partial \rho_n}}.$$

The thermodynamical spinodal region is then defined to be that region of $(\rho_p,\rho_n)$ space for which $\lambda_{-} < 0$. Matter in the spinodal region will separate into two phases: a low density gas phase and a higher density liquid phase. The higher density boundary of the spinodal gives an estimate for the transition density to uniform matter; given that the spinodal analysis neglects the competition between the Coulomb energy and surface energies of the two-phase system, that estimate is expected to provide a low limit on the transition density. By comparing the spinodal region with a fully microscopic calculation of the pasta phases up to the transition density, we will be able to evaluate the difference between that lower limit and the actual transition density.

\section{Computational method} \label{III}

As stated in section \ref{I}, we use a 3DSHF approximation with a phenomenological Skyrme model for the nuclear force. In the calculation, it is assumed that, at a given density and temperature, matter is arranged in a periodic structure throughout a sufficiently large region of space for a unit cell to be identified, in which the microscopic and bulk properties of the matter are calculated. The calculation is performed in cubic cells with periodic boundary conditions and assuming reflection symmetry across the three Cartesian axes. The required reflection symmetry allows us to obtain solutions only in one octant of the unit cell, which reduces significantly the computer time. The only effect of confining ourselves to $1/8$ of the cell is that we can only consider triaxial shapes.

It is expected that the absolute minimum of the free energy of a cell containing $A$ nucleons is not going to be particularly pronounced and there will be a host of local minima separated by relatively small energy differences. In order to systematically survey the ``shape space" of all nuclear configurations of interest, the quadrupole moment of the neutron density distributions has been parametrized, and those parameters constrained. It is expected that the proton distribution follows closely that of the neutrons.

The minimum of the free energy in a cell at a given particle number density, temperature and a proton fraction is sought as a function of 3 free parameters, the number of particles in the cell (determining the cell size) and the parameters of the quadrupole moment of the neutron distribution $\beta, \gamma$. Each minimization takes approximately 12 hours on a single CPU core in computers like the National Center for Computational Sciences (NCCS) Cray XT5/XK6 machine at ORNL and is performed in a trivially parallel mode, typically using 45,000 processors in one run to perform separate minimizations over a range of densities $0.02 - 0.12$ fm$^{-3}$, temperatures $0 - 10$ MeV and a fixed proton fraction of $y_p = 0.3$, where we have spent approximately $2.3 \times 10^6$ CPU hours. We shall refer generally to this implementation and computation as the 3DSHF model. Full computational details can be found in \citep{Newton-09, Newton-phd, Pais-12}.

\section{Numerical results and discussions} \label{IV}

We show in Table \ref{tab1} the nuclear matter properties for the four Skyrme models we use in this study, NRAPR, SQMC700, SkM$^*$ and SLy4. We chose two traditional forces, SkM* and SLy4, and two forces, NRAPR and SQMC700, which successfully satisfied up-to-date experimental and observational constraints on properties of nuclear matter \citep{Dutra-12}. These four forces span the range of saturation-density symmetry energy slope $L$ obtained from the concordance of a variety of experimental probes $40 \lesssim L \lesssim 60$MeV. We also show the nuclear matter properties of another model to which we will compare our results, the widely used Lattimer-Swesty (LS) EoS \citep{Lattimer-91}.

\begin{table}[!htbp]
  \centering
  \caption{Nuclear matter properties at saturation density $\rho_0$ (energy per particle $B/A$, incompressibility $K$, symmetry energy $E_{sym}$ and symmetry energy slope $L$) for the models studied. All the quantities are in MeV, except for $\rho_0$, given in fm$^{\rm -3}$.} 
  \begin{tabular}{c c c c c c}
    \hline
    \hline
	Model & $\rho_0$ & $B/A$ & $K$ & $E_{sym}$ & $L$   \\
    \hline
	NRAPR & 0.16 & -15.85 & 226 & 33 & 60 \\
	SQMC700 & 0.17 & -15.49 & 222 & 33 & 59  \\
	SkM* & 0.16 & -15.77 & 217 & 30 & 46 \\
	SLy4 & 0.16 & -15.97 & 230 & 32 & 46 \\
    LS & 0.155 & 16 & 220 & 29.3 & 74 \\
	\hline   
    \hline
  \end{tabular}
 \label{tab1}
\end{table} 

\begin{figure}
\begin{tabular}{c}
\includegraphics[width=0.45\textwidth]{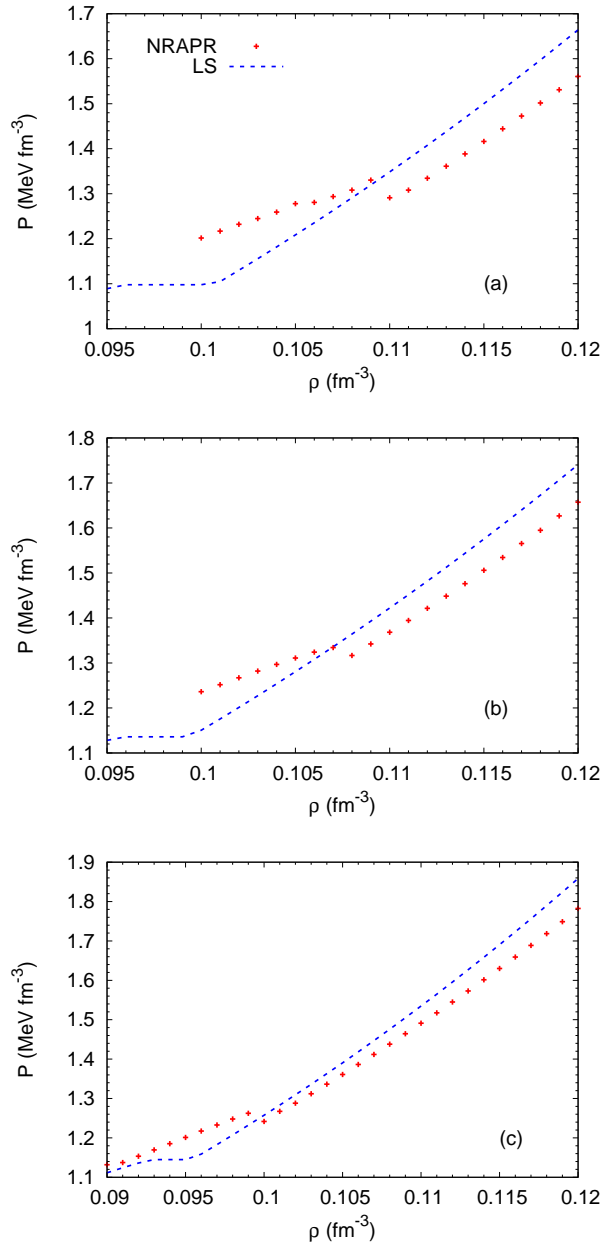} \\
\end{tabular}
\caption{(Color online) Pressure versus density for the NRAPR interaction (blue dashed line) and the Lattimer-Swesty EoS (red points) for (a) $T = 2$ MeV, (b) $T = 4$ MeV and (c) $T = 6$ MeV. The transition to uniform matter happens at different densities due to differences in the underlying nuclear interaction model used.} 
\label{fig1}
\end{figure}

When the EoS of supernova matter is assembled from separate treatments of the inhomogeneous and homogeneous phases, the Maxwell or Gibbs construction is needed to connect the two phases in a thermodynamically consistent way. In our 3DSHF model, the two phases are treated consistently with no need for such a construction. We contrast the results of our 3DSHF model using the NRAPR Skyrme parameterization with the LS EoS in Fig. \ref{fig1}, where the density dependence of pressure at $T = 2, 4, 6$ MeV is plotted.
 Since the LS EoS employs a Maxwell construction, they obtain a range of densities where the pressure is constant, removing, as a result, the pressure discontinuity that accompanies a first order phase transition. In our model, we see a clear pressure discontinuity at the density where the results of our simulations yield uniform matter. We thus obtain a clear indication of a first order phase transition. Note that the underlying nuclear interaction is different, so we should not expect an exact match for the uniform matter EoS. 

\begin{figure}
\begin{tabular}{c}
\includegraphics[width=0.5\textwidth]{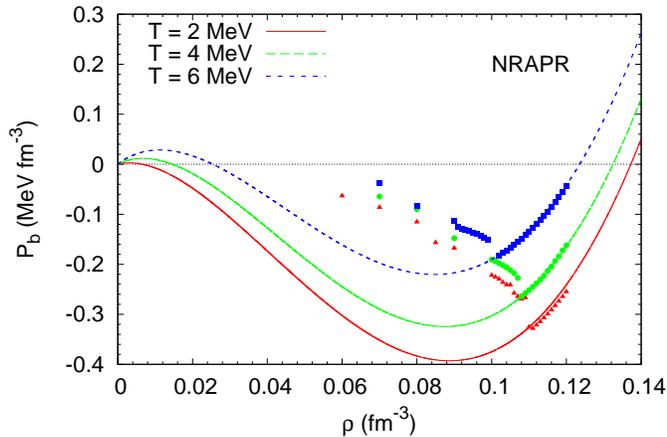} \\
\end{tabular}
\caption{(Color online) Baryonic pressure versus density for the NRAPR interaction and $T = 2$ MeV (red), $T = 4$ MeV (yellow) and $T = 6$ MeV (blue). The lines were calculated with the uniform matter code. The points with the 3DSHF code.} 
\label{fig2}
\end{figure}

In Fig. \ref{fig2}, we plot the baryonic pressure as a function of the density for the NRAPR model and some temperatures. We show the results obtained with the 3DSHF code (points). The jump in the pressure occurs when the transition to uniform matter happens. Of course, in the Skyrme-Hartree-Fock scheme, analytic expressions for the uniform matter EoS are obtained, and the EoS calculated exactly. A test of the accuracy of the 3DSHF code is provided by comparing its predictions for uniform matter EoS against the analytic result, shown as the continuous lines. In practice, owing to the finite computational volume of the 3DSHF implementation, spurious shell effects arising from the discretization of single particle states prevent an exact match. Here we calibrate the 3DSHF model by adding a small correction factor to the predicted thermodynamical quantities such that agreement with the analytic results is obtained. All results shown include this correction factor.

\begin{figure}
   \begin{tabular}{c c}
   \includegraphics[width=0.45\textwidth]{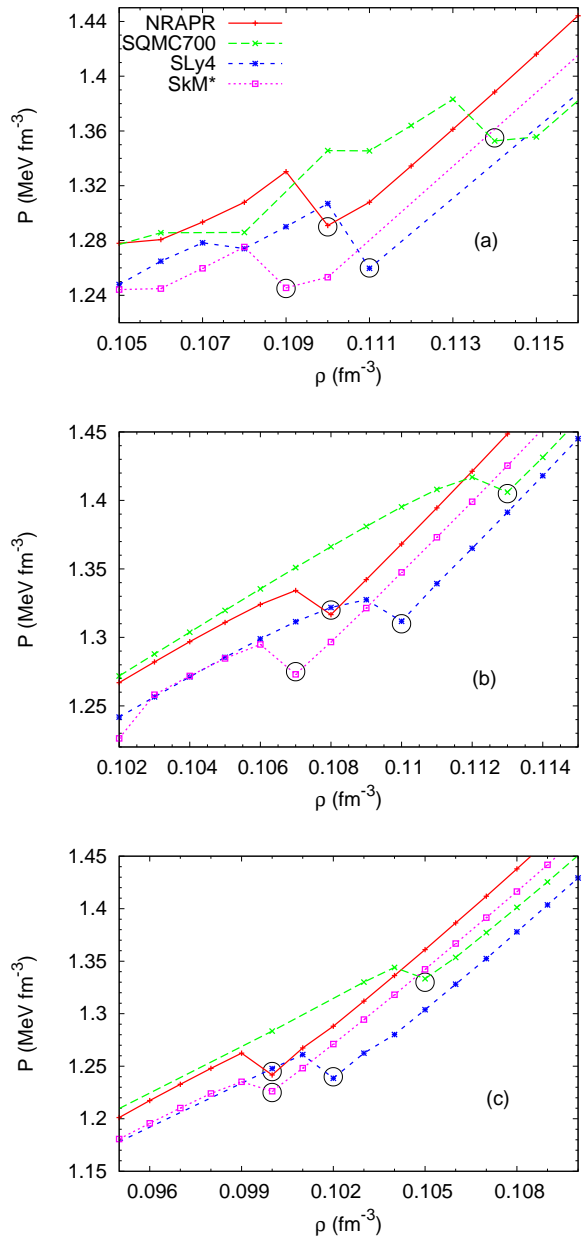} \\
   \end{tabular}
\caption{(Color online) Pressure as a function of the number density for the Skyrme interactions for (a) $T = 2$ MeV, (b) $T = 4$ MeV and (c) $T = 6$ MeV. The points circled are the onset densities of homogeneous matter (see Ref. \citep{Pais-12}).}%
\label{fig3}
\end{figure}

\begin{figure}
   \begin{tabular}{c c}
   \includegraphics[width=0.45\textwidth]{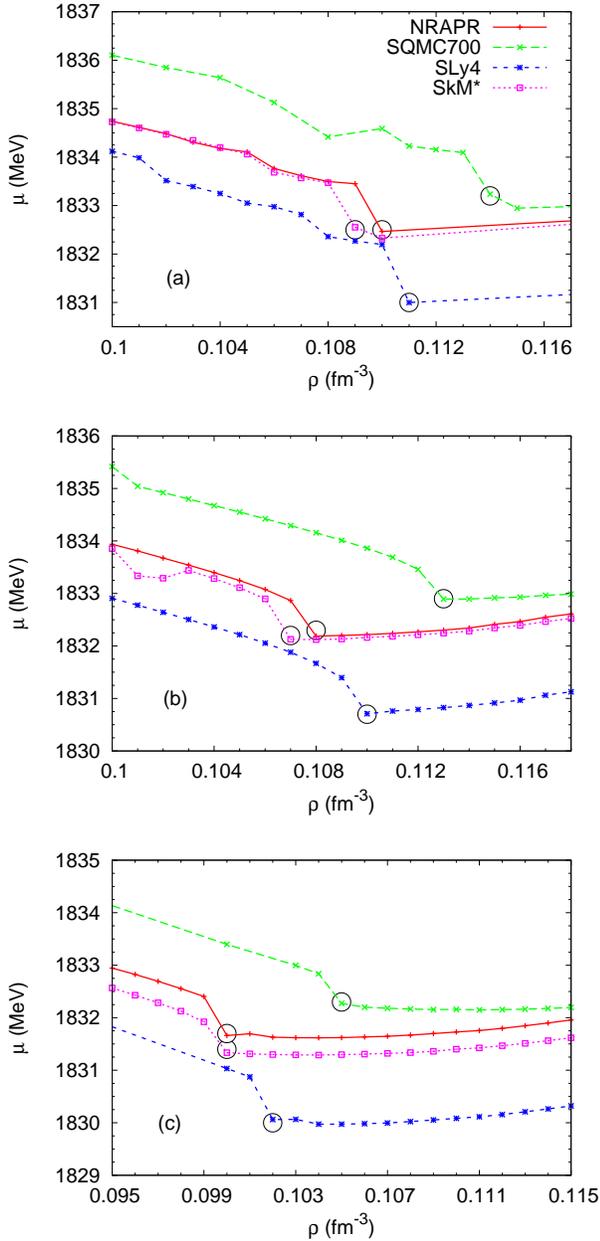} \\
   \end{tabular}
\caption{(Color online) Chemical potential as a function of the number density for the Skyrme interactions for (a) $T = 2$ MeV, (b) $T = 4$ MeV and (c) $T = 6$ MeV. The points circled are the onset densities of homogeneous matter (see Ref. \citep{Pais-12}).}%
\label{fig4}
\end{figure}

\begin{figure}
   \begin{tabular}{c}
   \includegraphics[width=0.45\textwidth]{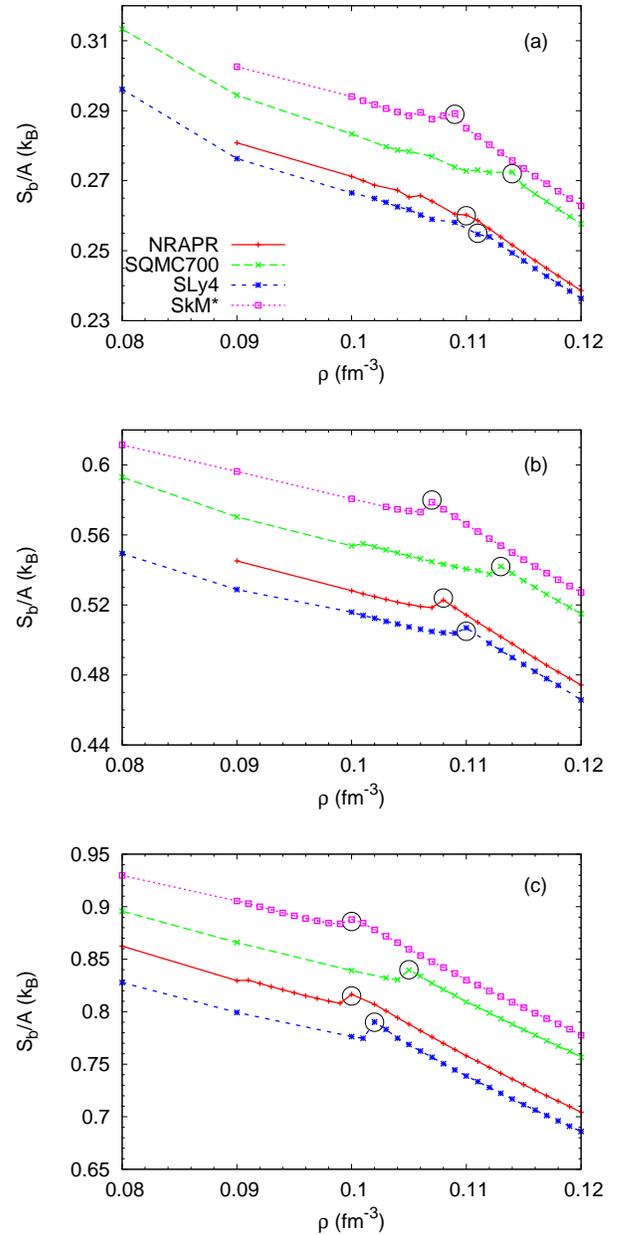}
   \end{tabular}
\caption{(Color online) Baryonic entropy per particle as a function of the density. All the Skyrme interactions are shown. The range of temperatures is (a) $T = 2$ MeV, (b) $T = 4$ MeV, and (c) $T = 6$ MeV. The points circled are the onset densities of homogeneous matter (see Ref. \citep{Pais-12}).}
\label{fig5}
\end{figure}

To further explore the fingerprint of the first-order phase transition on the predicted thermodynamic quantities, we plot as a function of density the pressure (Fig. \ref{fig3}), baryonic chemical potential (Fig. \ref{fig4}), and the baryonic entropy per baryon (Fig. \ref{fig5}) at temperatures of 2, 4 and 6 MeV. In all cases the transition to uniform matter is highlighted  by a circle, and the pressure, chemical potentials and entropies exhibit the characteristic discontinuity of a first-order phase transition. In Fig. \ref{fig5}, the baryonic entropy per particle decreases with the density.

These results are in qualitative agreement with statistical models, see, e.g. Fig. 20 of Ref. \citep{Raduta10}, except for the important difference that in our calculation, we obtain a discontinuity in the entropies, pressure, chemical potentials with respect to density, indicating the first-order phase transition, whereas in, for example, Ref. \citep{Raduta10}, their results vary continuously with the density (see Figs. 12, 20 and 23 of Ref. \citep{Raduta10}), thus obtaining a continuous phase transition. In such a statistical model, matter at sub-saturation densities is modeled as a continuous fluid mixture between free nucleons and massive nuclei. Their model does not take into account the pasta phases, however. Within our model, we obtain unambiguously a first-order phase transition.

In Figs. \ref{fig3}, \ref{fig4}, \ref{fig5}, we also see small jumps in the thermodynamic properties at lower temperatures between different phases of pasta. Williams et al. \cite{Williams85} found first-order phase transitions between the various pasta phases and to uniform matter marked by discontinuities in the pressure and chemical potential. In our model, the discontinuities in the first derivatives of the free energy density at lower densities are too small to be unambiguously identified. These jumps are smoothed out at higher temperatures, pointing to an origin in quantum shell effects. 
In addition, the excitation energies of pasta structures are of order MeV \cite{Grasso-08}; a statistical model taking into account such excitations is beyond the scope of the current work, but might lead
to additional smoothing of the thermodynamic quantities through the densities at which the pasta phases exist.

\subsection{Comparison with thermodynamic spinodal}

\begin{figure*}
   \begin{tabular}{c c}
   \includegraphics[width=\textwidth]{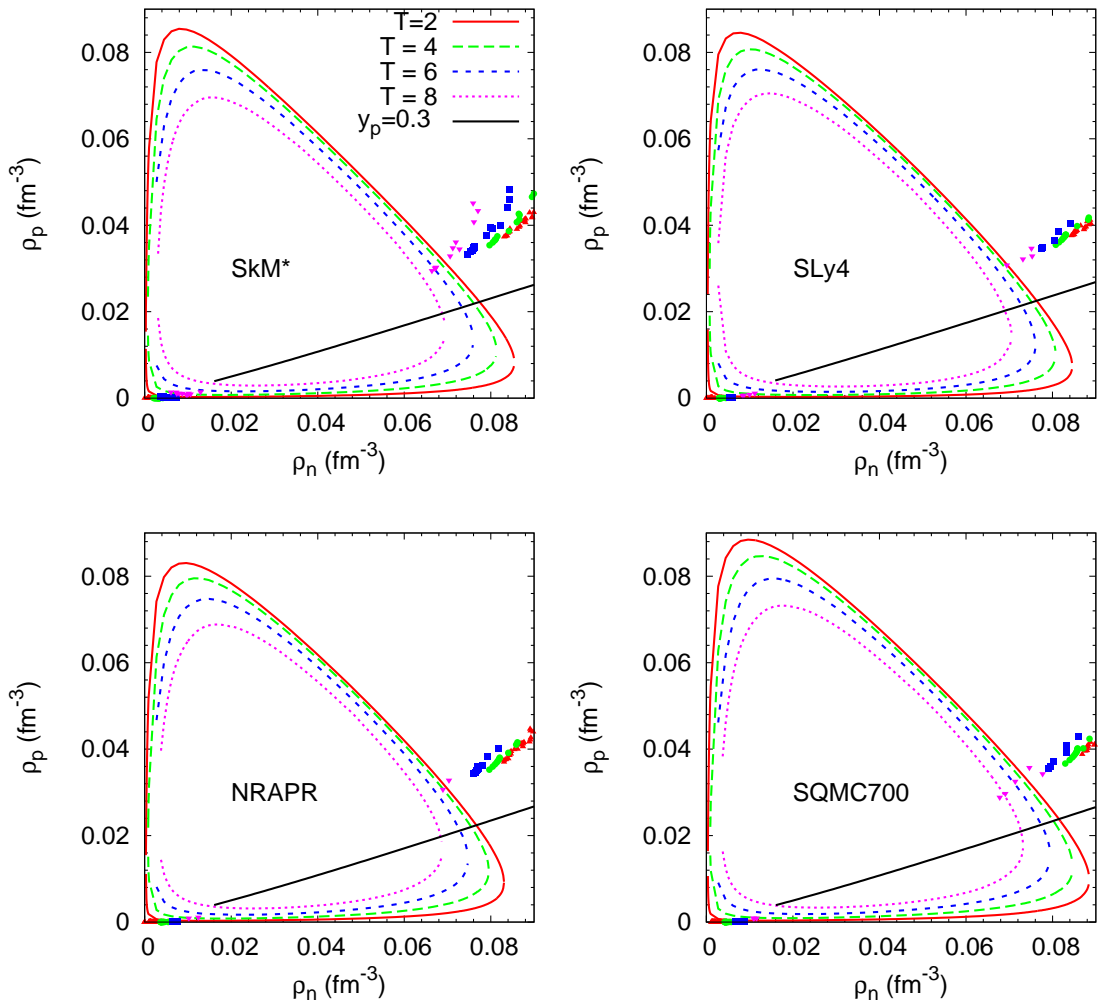} \\
   \end{tabular}
\caption{(Color online) Thermodynamical spinodals for the Skyrme interactions, and $2\leq T \leq 8$ MeV. The black line that crosses the spinodals is the EOS for uniform matter with $y_p = 0.3$. The points shown in the spinodals are the maximum (liquid phase) and minimum (gas phase) values for the neutron density and the correspondent proton density, for each equilibrium configuration obtained with the 3DSHF code.}%
\label{fig6}
\end{figure*}

In Fig. \ref{fig6} we plot the thermodynamic spinodals for the Skyrme interactions, and a range of temperatures $T = 2-8$ MeV. Inside the spinodals, matter is unstable to density and composition fluctuations, and is predicted to decompose into coexisting gas and liquid phases; physically, this is expected to result in the nuclear pasta phases. The unstable region decreases as temperature increases \citep{Pais10}. In order to estimate the transition density to uniform matter, we need to add to these thermodynamical spinodals the equation of state. The lines shown are for a proton fraction of 0.3. One can estimate the transition density to uniform matter by taking it to be the point at which the $y_p = 0.3$ EOS crosses the spinodal on the high density side (these transition values will later be labelled as TS). Then non-homogeneous phases in supernova matter correspond to the EOS inside the spinodal \citep{Pais09}. For matter with a proton faction of 0.3 (the case for CCSN matter), the nonhomogeneous phase will still exist for $T = 10$ MeV. Within the range of EoS parameters considered here, motivated by experimental constraints, the spinodals are almost coincident except for SQMC700  and the transition densities to uniform matter can be taken with confidence to be well defined, and are given in Table~\ref{tab2}. Given that the main difference between the SQMC700 and the other Skyrme parameterizations is the higher saturation density, we can conclude that variations of the transition density with respect to the symmetry energy within its uncertain range at saturation density is small.

Neglecting short-range correlations, if uniform matter finds itself at such a density and composition such that it lies inside the spinodal, it will undergo phase separation into a low density gas phase, lying to the left of the spinodal, and a high density liquid phase, to the right of the spinodal. Physically, the liquid phase would correspond to the pasta phases. In actuality, the effect of short-range Coulomb and surface correlations stabilize the pasta at higher densities as manifest in the transition densities discussed above. To probe this further, we estimate the proton and neutron densities inside and outside the pasta phases that result from our 3DSHF calculations by simply taking the highest proton/neutron density found in the unit cell to represent the density in the liquid phase, and the lowest density to represent the gas phase. These obviously constitute an upper limit to the liquid density and a lower limit to the gas density. We plot these points in Fig. \ref{fig6} overlaid on the spinodals. 

In the simple coexisting phases picture, one would see the densities in the liquid phases reach down to the upper spinodal boundary. Noticeably for temperatures up to $T=6-8$MeV, the liquid phase in the 3DSHF calculations exist at densities higher by around $0.01$fm$^{-3}$ than the upper spinodal boundary. This means the effective boundary between the uniform matter phase and past phases is larger by $\sim0.01$fm$^{-3}$ compared to the simple thermodynamic spinodal picture, leading to the higher transition densities found above. This is in broad agreement with the results obtained using the Thomas-Fermi formalism \citep{Avancini-12}.

\begin{table}
  \centering
  \caption{Onset density of homogeneous matter for the four models and temperatures considered. For more explanations see text.}  
  \begin{tabular}{c|cc|cc|cc|cc}
    \hline
    \hline
Model     & \multicolumn{2}{c}{ NRAPR } & \multicolumn{2}{c}{ SQMC700 } & \multicolumn{2}{c}{ SkM* } & \multicolumn{2}{c}{ SLy4 } \\
$T$ [MeV] & \multicolumn{8}{c}{ $\rho_{trans}$ (fm$^{-3}$)}   \\
\hline
	      &	3D    & TS  &	3D    &  TS  &	3D   &  TS  & 3D     & TS  \\	
 2        & 0.110 & 0.099 & 0.114  & 0.105 & 0.109  & 0.100	& 0.111  & 0.099 \\
 4        & 0.108 & 0.097 & 0.113  & 0.103 & 0.107  & 0.098	& 0.110  & 0.097  \\
 6        & 0.100 & 0.094 & 0.105  & 0.100 & 0.100  & 0.094	& 0.102  & 0.094  \\
 8        & 0.091 & 0.088 & 0.099  & 0.094 & 0.089  & 0.088	& 0.094  & 0.089  \\
 \hline\hline
  \end{tabular}
\label{tab2}
\end{table}

\begin{figure*}
   \begin{tabular}{c c}
   \includegraphics[width=\textwidth]{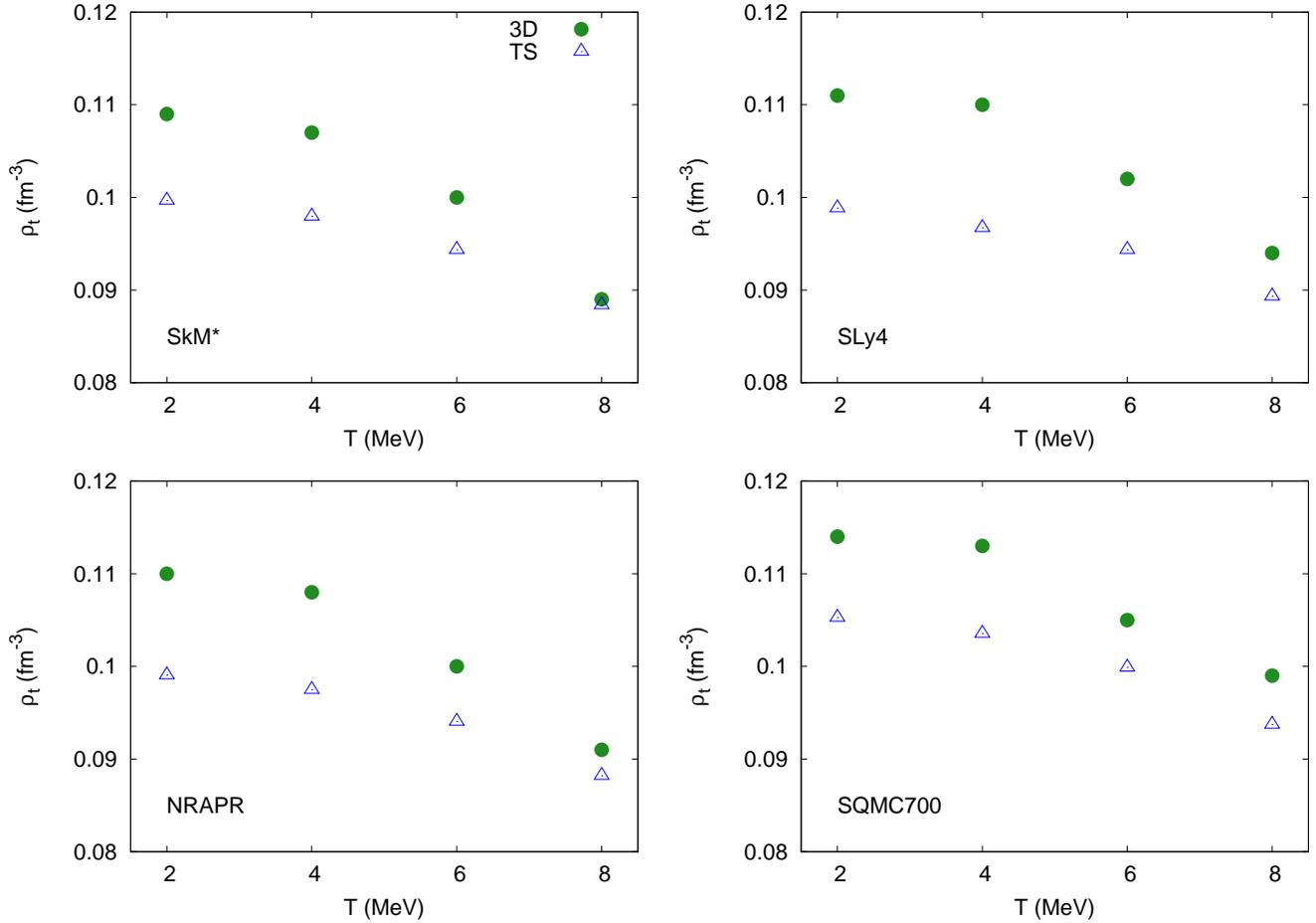} \\
   \end{tabular}
\caption{(Color online) Transition densities as a function of the temperature, for $y_p = 0.3$, and all the interactions considered. The points correspond to the values shown in Table \ref{tab2}.}%
\label{fig7}
\end{figure*}

The transition densities estimated by the TS calculation should constitute a lower limit on the transition density \citep{Pethick-95}. In Table~\ref{tab2} and Fig.~\ref{fig7} we compare the transition densities obtained from the spinodal analysis with those from the 3DSHF calculations, and find that for temperatures $T\lesssim10$MeV that indeed the spinodal transition densities are lower (by $\sim0.01$fm$^{-3}$ for temperatures $T\lesssim5$MeV.) At higher temperatures, the difference decreases as temperature effects come to dominate over the short-range Coulomb and surface energies which lend stability to matter at lower temperatures. 

\section{Conclusions} \label{V}

A self-consistent 3D-Hartree-Fock calculation which includes Coulomb, surface and quantum finite size effects and allows for an unbiased exploration of all possible triaxial nuclear pasta structures has been performed for proton fractions $y_{\rm p}=0.3$ and temperatures $T=2-8$ MeV relevant to core-collapse supernovae. Using this model, we have studied the transitions between the pasta phases and to uniform matter.

We see a clear indication of a first order phase transition to uniform matter, manifesting itself as discontinuities in thermodynamic quantities such as the pressure, entropy and chemical potentials. We were able to identify jumps (of the order of $\mu_{\rm b} = 1$ MeV) in all the first derivatives of the free energy. The inclusion of the pasta phase in our model hence does not remove the first order phase transition to uniform matter. Discontinuities in the first derivatives of the free energy density at lower densities are also present, indicating possible small first-order transitions between the pasta phases, but are too small to be identified unambiguously owing to the numerical accuracy of the computational framework. However we can observe that they happen gradually, with increasing density, and that first-order phase transitions between the pasta phases are likely weak at most. 

We find that the current range of uncertainty in the symmetry energy at saturation density has only a small effect on the transition density to uniform matter, which otherwise can be taken to be well-determined. The difference between the transition densities for all Skyrmes, at all temperatures considered here, is around 0.005 fm$^{-3}$. However, a comparison of the results of our 3DSHF calculations with the spinodal analysis leads us to conclude that short-range Coulomb correlations and quantum shell effects stabilize structures that would otherwise be unstable. This leads to a modification in the transition density compared with that obtained using the spinodals, with the spinodal method underestimating the densities by up to $\sim 0.01$fm$^{-3}$ at the lowest temperatures. This difference becomes progressively smaller at higher temperatures, becoming negligible at $T\sim8$MeV.

It is possible that degrees of freedom not taken into count in the mean-field 3DSHF calculations can modify our conclusions. Particularly, the fact that we see a decrease in the chemical potential with the density, indicates that our description of the pasta phase, with heavy clusters and a background gas, may be missing some degrees of freedom that will maximize the entropy (see \citep{Raduta10}), and that such a continuous mixture may soften the first-order phase transition we observe. 

The results of this work, extended to relevant temperatures and proton/neutron ratios, can be used to construct four EoS for supernova simulation models, augmented by 1D calculation at densities below and above the pasta region. Neutron and proton density distributions in the unit cell, obtained in this work, can also be employed in the modeling  of neutrino transport through the pasta formations. At low momentum transfers the static structure factor is found to be small because of ion screening. In contrast, at intermediate momentum transfers the static structure factor displays a large peak due to coherent scattering from all the neutrons in a cluster. This peak moves to higher momentum transfers and decreases in amplitude as the density increases \cite{Horowitz-04II}. A large static structure factor at zero momentum transfer, indicative of large density fluctuations during a first-order phase transition, may increase the neutrino opacity. 

\section*{ACKNOWLEDGMENTS}

We are very grateful to C. Provid\^encia for useful discussions.
This research used resources of the Oak Ridge Leadership Computing Facility, located in
the National Center for Computational Sciences at Oak Ridge National Laboratory, which is
supported by the Office of Science of the Department of Energy under Contract DE-AC05-00OR22725. It was supported by an INCITE grant AST005: Multidimensional Simulations of Core-Collapse Supernovae, PI: Anthony Mezzacappa, Oak Ridge National Laboratory. Partial support comes from ``NewCompStar'', COST Action MP1304. H.P. is supported by FCT under Project No. SFRH/BPD/95566/2013. She also thanks the INFN for partial support and the Galileo Galilei Institute for Theoretical Physics, where parts of this work were carried out, for the hospitality. W.G.N. was supported by the National Aeronautics and Space Administration under grant NNX11AC41G issued through the Science Mission Directorate.

\end{document}